\documentstyle[axodraw]{frank}
\def\beq{\begin{equation}}
\def\eeq{\end{equation}}
\def\bea{\begin{eqnarray}}
\def\eea{\end{eqnarray}}
\def\barr{\begin{array}}
\def\earr{\end{array}}
\newcommand{\sm}{standard model}
\newcommand{\cm}{center of mass}
\newcommand{\xs}{cross section}
\newcommand{\lc}{linear collider}
\newcommand{\pe}{\mbox{$e^+e^-$}}
\newcommand{\ee}{\mbox{$e^-e^-$}}

\begin{document}

\begin{flushright}
PSI-PR-96-09\\
February 1996
\end{flushright}

\vskip2cm

\begin{frontmatter}
\title{Comparative Study of $Z'$ Searches \\
in $e^+e^-$ and $e^-e^-$ Scattering}
\author{Frank Cuypers}
\address{{\tt cuypers@pss058.psi.ch}\\
        Paul Scherrer Institute,
        CH-5232 Villigen PSI,
        Switzerland}
\begin{abstract}
We consider indirect searches 
for additional neutral vector bosons
in $e^+e^-$ and $e^-e^-$ collisions,
and compare these two \lc\ modes 
with similar analysis procedures and assumptions.
Discovery limits and resolving power 
are discussed
in a model-independent way.
\end{abstract}
\end{frontmatter}
%%\journal{Physics Letters B}
%%\journal{Nuclear Physics B}
\clearpage

\section{Introduction}

In contrast to storage rings,
it is straightforward to replace a positron beam by an electron beam
at a \lc.
There are several important characteristics
which differentiate $e^-e^-$ from $e^+e^-$ collisions~\cite{e-e-}
and justify considering both options on the same footing
when it comes to designing the \lc s of the next generation,
generically the Future Linear Collider (FLC):
\begin{itemize}
\item
  The $e^-e^-$ environment is much cleaner
  because there is much less \sm\ activity.
  In particular,
  since QCD enters the game only at much higher orders
  and is associated with missing energy,
  the systematic errors 
  due to the possible misidentification of electrons and pions
  are negligible.
\item
  Current electron guns
  can already produce beams with polarizations exceeding 75\%.
  There is no doubt that even further improvements are due.
  It is not clear, however, 
  whether polarized positron beams may ever be obtained at all.
  The $e^-e^-$ collisions therefore
  offer the possibility of polarizing both initial states
  and to perform three independent experiments.
\item
  The $e^-e^-$ initial state
  is not only doubly charged,
  but also carries a finite lepton number.
  This allows to explore areas of new physics
  which are, 
  at best,
  difficult to access in conventional $e^+e^-$ annihilations.
\end{itemize}

\noindent
We investigate here how a $Z'$ can be discovered and studied
in $e^\pm e^-$ scattering.
For the $e^+e^-$ reaction
we concentrate on muon pair production~\cite{l}
which provides a very clean signal,
whereas for the $e^-e^-$ process
we analyze corrections to M\o ller scattering~\cite{ccl}.

The next section is devoted to a short description of some models
predicting the existence of a $Z'$.
After this,
we discuss the expected $e^\pm e^-$ \xs s.
We then proceed to a comparison of the discovery potential of each reaction,
both at moderate SLC and LEP2 energies,
as well as at the design FLC energy of 500 GeV.
In the event a $Z'$ is discovered,
we examine how precisely the three parameters involved
(vector and axial couplings and mass)
can be measured
and compare these expectations
with the predictions of some models. 
Finally,
we shortly comment on the incidence of radiative corrections.

\section{\boldmath $Z'$ Models}

There are many extensions of the \sm\ 
which predict the existence of extra neutral vector bosons $Z'$,
and most searches are performed within the framework 
of a particular model~\cite{a}.
It was advocated, though,
that it is important to perform a model-independent analysis~\cite{l},
in order not to miss out on some unexpected new kind of physics
lurking beyond the \sm.
We also adopt this point of view here.

Assuming lepton universality,
the generic lagrangian describing the interaction
of a heavy neutral vector boson $Z'$ with the charged leptons
can be written
\beq
L = e~\bar\psi_\ell \gamma^\mu \left( {v_{Z'}} + {a_{Z'}}\gamma_5 \right) \psi_\ell ~{Z'_\mu}
\label{lag}~,
\eeq
where $v_{Z'}$ and $a_{Z'}$ are the vector and axial couplings
normalized to the charge of the electron $e$.
This interaction
mediates both $e^+e^-$ annihilation
and $e^-e^-$ scattering.
The results of LEP1 constrain any $Z^0$-$Z'$ mixing
to such an extent \cite{ll}
that it can safely be ignored here.

In a model-independent analysis
the two couplings in the lagrangian (\ref{lag}) are free.
In the realm of specific models, 
however,
their values are constrained or even fixed.
We summarize now a few commonly used models,
and will refer to these particular choices later,
when it comes to gauge the capabilities of a \lc\
operated in both $e^+e^-$ and $e^-e^-$ modes.
Note that for all these models
the $Z'$ mass remains a free parameter.
The trajectories or locations 
in the $\left(v_{Z'},a_{Z'}\right)$ plane
of these models 
are shown in Fig.~\ref{fmodels}

\subsection{Sequential Standard Model}
In this model
it is assumed that there is a heavy $Z'$ vector boson
which couples to the fermions 
in the same way as the $Z^0$:
\beq
SSM: \quad \left\{
\begin{array}{rclcl}
v_{Z'} & = & v_{Z^0} & = & \displaystyle{1-4\sin^2\theta_w \over 4\sin\theta_w\cos\theta_w}

\\
a_{Z'} & = & a_{Z^0} & = & \displaystyle{-1 \over 4\sin\theta_w\cos\theta_w}~.
\end{array}
\right.
\label{ssm}
\eeq
This model is neither realistic nor even gauge invariant,
but some authors treat it as a useful benchmark.

\subsection{Un-unified Standard Model}
This model \cite{gjs} assumes leptons and quarks 
transform according to different $SU(2)_L$ weak groups.
The resulting $Z'$ interacts then only with left handed fermions,
and its couplings to the charged leptons are:
\beq
UU: \quad \left\{
\begin{array}{rcl}
v_{Z'} & = & ~~\displaystyle{1\over2\sin\theta_w} \quad \tan\phi
\\
a_{Z'} & = &  -\displaystyle{1\over2\sin\theta_w} \quad \tan\phi
\end{array}
\right.
\qquad\qquad
.22 \leq \sin\phi \leq .99 ~,
\label{uusm}
\eeq
where $\theta_w$ is the weak mixing angle.

\subsection{Left-Right Symmetric Models}
These models are based on the gauge group
$SU(2)_L \otimes SU(2)_R \otimes U(1)$.
This group can emerge as an intermediate symmetry
from the breaking of a $SO(10)$ grand unified theory,
in which case the $Z'$ couplings to the charged leptons are \cite{rabi}:
\beq
LR: \quad \left\{
\begin{array}{rcl}
v_{Z'} & = & ~~\displaystyle{1\over4\cos\theta_w} \quad \displaystyle{2-\beta^2\over\beta}
\\
a_{Z'} & = &  -\displaystyle{1\over4\cos\theta_w} \quad \beta
\end{array}
\right.
\qquad
\left\{
\begin{array}{l}
\beta = \sqrt{\displaystyle{\kappa^2\over\tan^2\theta_w}-1} > 0
\\
\kappa = \displaystyle{g_R\over g_L} > \tan\theta_w
\end{array}
\right.
\label{lrsm}
\eeq
where $g_{L,R}$ are the couplings constants of the two $SU(2)$ groups.
If these two couplings are equal,
one talks of the manifestly left-right symmetric model
(MLR).

In the so-called alternative left-right symmetric model
the $SU(2)_L \otimes SU(2)_R \otimes U(1)$ group
originates as a subroup of $E_6$ \cite{bunch}.
The couplings to the charged leptons
are then related to the above ones by the simple replacement
$\beta \leftrightarrow -1/\beta$
and read
\beq
ALR: \quad \left\{
\begin{array}{rcl}
v_{Z'} & = & \displaystyle{1\over4\cos\theta_w} \quad \displaystyle{1-2\beta^2\over\beta}
\\
a_{Z'} & = &  \displaystyle{1\over4\cos\theta_w} \quad \displaystyle{1\over\beta}
\end{array}
\right.
\label{alrsm}
\eeq
Again,
we define also the alternative manifestly left-right symmetric model
(AMLR)
by assuming
$g_L=g_R$.

\subsection{Grand Unified Theories}
The linear combination of two independent $U(1)$ subgroups
of a grand unified $E_6$ gauge group
yield the following type of $Z'$ couplings to the charged leptons \cite{gut}:
\beq
GUT: \quad \left\{
\begin{array}{rcl}
v_{Z'} & = & ~~~\displaystyle{1\over\sqrt{6}\cos\theta_w} \quad \cos\beta
\\
a_{Z'} & = &   -\displaystyle{1\over2\sqrt{6}\cos\theta_w} %\quad 
\left( \cos\beta + \sqrt{\displaystyle{5\over3}} \sin\beta \right)
\end{array}
\right.
\qquad
-{\pi} \leq \beta \leq {\pi} ~.
\label{gut}
\eeq
There are some favoured values of the mixing angle $\beta$,
which correspond to particular symmetry breaking chains.
They are labeled $\psi$, $\chi$ and $\eta$:
\beq
\psi: \quad \beta={\pi\over2} 
\qquad\qquad \left( E_6 \to SO(10) \otimes U(1)_\psi \right)
\label{psi}
\eeq
\beq
\chi: \quad \beta=0 
\qquad\qquad \left( SO(10) \to SU(5) \otimes U(1)_\chi \right)
\label{chi}
\eeq
\beq
\eta: \quad \tan\beta=-\sqrt{5\over3} 
\qquad \left( \mbox{superstring inspired} \right)~.
\label{eta}
\eeq

\section{Cross Sections}

The Born level Feynman diagrams for the
$e^+e^- \to \mu^+\mu^-$ and
$e^-e^- \to e^-e^-$ reactions
are shown in Figs~\ref{feyn}.
Neglecting the electron mass,
the differential \xs s are
\bea
\label{pe}
{d\sigma^{e^+e^-}\over d\cos\theta} & = &
{\pi\alpha^2 \over 4s} \quad
\displaystyle\sum_{i,j} \quad
{1 \over \left(\strut 1-y_i\right) \left(\strut 1-y_j\right)} \quad
\\\nonumber
&&\hspace{0em}
\biggl\{ ~
{1-P\over2} \quad L_iL_j
\left[~
  L_iL_j (1-\cos\theta)^2 ~+~ R_iR_j (1+\cos\theta)^2 
~\right]
\\\nonumber
&&\hspace{0em}
+ ~
{1+P\over2} \quad R_iR_j
\left[~
  R_iR_j (1-\cos\theta)^2 ~+~ L_iL_j (1+\cos\theta)^2 
~\right]
~ \biggr\}
\\\nonumber\\\nonumber\\
\label{ee}
{d\sigma^{e^-e^-}\over d\cos\theta} & = &
{2\pi\alpha^2 \over s} \quad
\displaystyle\sum_{i,j} \quad
{1 \over \left(\strut x_i^2-\cos^2\theta\right) \left(\strut x_j^2-\cos^2\theta\right)} \quad
\\\nonumber
&&\hspace{-6em}
\biggl\{ ~
{1-P_1-P_2+P_1P_2\over4} \quad L_i^2L_j^2 \quad 8x_i x_j
\\\nonumber
&&\hspace{-6em}
+ ~
%\qquad + \qquad
{1+P_1+P_2+P_1P_2\over4} \quad R_i^2R_j^2 \quad 8x_i x_j
\\\nonumber
&&\hspace{-6em}
+ ~
{1-P_1P_2\over2} \quad R_iL_iR_jL_j \quad
  \left[
    x_i x_j + (1+x_i x_j+2x_i+2x_j)\cos^2\theta + \cos^4\theta
  \right]
~ \biggr\}
\eea

\medskip
\beq
y_i = {m_i^2 \over s}
\qquad\qquad
x_i = 1 + 2y_i
\qquad\qquad
i,j=\gamma,{Z^0},Z'
\eeq

\beq
\left\{
  \barr{l}
  \strut R_i = v_i + a_i \\
  \strut L_i = v_i - a_i \makebox(0,20)[b]{}
  \earr
\right.
\qquad
\left\{
  \begin{array}{l}
  \strut v_\gamma = 1 \\
  \strut a_\gamma = 0
  \earr
\right.
\qquad
\left\{
  \begin{array}{l}
  \strut v_{Z^0} = \displaystyle{1-4\sin^2\theta_w \over 4\sin\theta_w\cos\theta_w} \\
  \strut a_{Z^0} = \displaystyle{-1 \over 4\sin\theta_w\cos\theta_w}~.
  \earr
\right.
\eeq
The polarizations of the electron beams are given by $P$ and $(P_1,P_2)$,
in the \pe\ (\ref{pe}) and the \ee\ (\ref{ee}) reactions respectively.
In our convention,
a fully right(left) polarized beam is given by $P=1(-1)$.

We see that the $e^+e^-$ \xs\ (\ref{pe})
is asymmetric,
whereas
the $e^-e^-$ \xs\ (\ref{ee}) 
is symmetric
in the scattering angle.
The resulting total \xs s
in the presence and absence of a $Z'$
are shown for unpolarized beams
in Fig.~\ref{fxs},
where we imposed an angular cut of $10^o$ 
on the scattering angle.
The total \xs\ of the $e^-e^-$ reaction
is very insensitive to the presence of a heavy $Z'$,
whereas the $e^+e^-$ \xs\ is strongly influenced 
by the tail of the $Z'$ resonance.

It is instructive to estimate these \xs s
in the limit where
\beq
m_{Z^0}^2 \ll s \ll m_{Z'}^2
\qquad \mbox{and} \qquad
\sin^2\theta_w = {1\over4}
\label{approx}~.
\eeq
With these approximations we have
\beq
\left\{
  \begin{array}{l}
  y_\gamma = y_{Z^0} = 0 \\
  y_{Z'} = m_{Z'}^2/s \makebox(0,20)[b]{} \\
  x_\gamma = x_{Z^0} = 1 \makebox(0,20)[b]{} \\
  x_{Z'} = 2m_{Z'}^2/s \makebox(0,20)[b]{}
  \earr
\right.
\qquad
\left\{
  \begin{array}{l}
  v_e = 1 \\
  a_e = 0 \makebox(0,20)[b]{} \\
  v_{Z^0} = 0 \makebox(0,20)[b]{} \\
  a_{Z^0} = -1/\sqrt{3} \makebox(0,20)[b]{}
  \earr
\right.
\qquad
\left\{
  \begin{array}{l}
  R_e = 1 \\
  L_e = 1 \makebox(0,20)[b]{} \\
  R_{Z^0} = -1/\sqrt{3} \makebox(0,20)[b]{} \\
  L_{Z^0} = 1/\sqrt{3} \makebox(0,20)[b]{}~.
  \earr
\right.
\eeq
It is then easy to compute the \sm\ contributions,
who represent the bulk of the \xs:
\bea
{d\sigma^{e^+e^-}\over d\cos\theta} & = &
{\pi\alpha^2 \over 9s} %\quad
\left( 5 - 6\cos\theta + 5\cos^2\theta \right)
\label{pebulk}
\\\nonumber\\
{d\sigma^{e^-e^-}\over d\cos\theta} & = &
{8\pi\alpha^2 \over 9s} %\quad
{1 \over \sin^4\theta}
\biggl[~
{1+P_1P_2 \over 2} ~ 32
\\
&&\nonumber\qquad\qquad\quad
+~ {1-P_1P_2 \over 2} ~ \left( 1 + 6\cos^2\theta + \cos^4\theta \right) 
~\biggr]
~.
\label{eebulk}
\eea

The $e^-e^-$ \xs\ 
has thus always a strong polarization dependence.
In contrast,
the $e^+e^-$ \xs\ 
should thus not depend very much on the electron beam polarization.

\section{Discovery Limits}

For small $Z'$ couplings or 
large $Z'$ masses,
the signal will mainly be confined in the interference terms.
In the limits (\ref{approx})
we find
\bea
\Delta{d\sigma^{e^+e^-}\over d\cos\theta} & = &
{\pi\alpha^2 \over 3m_{Z'}^2} \quad
\biggl\{ ~
{1-P\over2} ~ L_{Z'}
\left[~
  2L_{Z'}(1-\cos\theta)^2 ~+~ R_{Z'}(1+\cos\theta)^2 
~\right]
\nonumber\\\label{pesignal}
&&\hspace{1em}
+ ~
{1+P\over2} ~ R_{Z'}
\left[~
  2R_{Z'}(1-\cos\theta)^2 ~+~ L_{Z'}(1+\cos\theta)^2 
~\right]
~ \biggr\}
\\\nonumber\\\nonumber\\
\Delta{d\sigma^{e^-e^-}\over d\cos\theta} & = &
{4\pi\alpha^2 \over 3m_{Z'}^2} \quad
{1 \over \sin^2\theta} \quad
\Biggl[~
  16 ~R_{Z'}^2~ {1+P_1+P_2+P_1P_2 \over 4} 
\nonumber\\\label{eesignal}&&\hspace{6.5em}
~+~  16 ~L_{Z'}^2~ {1-P_1-P_2+P_1P_2 \over 4} 
\\\nonumber&&\hspace{6.5em}
~+~  R_{Z'}L_{Z'}~ {1-P_1P_2 \over 2} ~(1+4\cos^2\theta)
~\Biggr]
~.
\eea

With fully polarized beams
the $e_R^-e_R^-$ ($P_1=P_2=1$)
and $e_L^-e_L^-$ ($P_1=P_2=-1$) 
reactions can probe independently
$R_{Z'}$ and $L_{Z'}$.
The third term in Eq.~(\ref{eesignal})
is strongly suppressed
and a measurement in the $e^-_Le^-_R$ ($P_1=-P_2=1$) mode 
will therefore not provide very much information.
For this reason
we shall ignore this combination of polarizations for the time being.

We can combine Eqs~(\ref{pesignal},\ref{eesignal}) 
with Eqs~(\ref{pebulk},\ref{eebulk}) 
to derive the Cramer-Rao bound 
on the smallest values of the $Z'$ couplings 
one can observe in each experiment~\cite{cr}:
\beq
\chi_\infty^2 = {\cal L} \int_{-1}^{+1} d\cos\theta ~
{ 
  \left( \Delta\displaystyle{d\sigma\over d\cos\theta} \right)^2
  \over
  \displaystyle{d\sigma\over d\cos\theta}
}
\label{cr}~.
\eeq
Assuming both experiments are performed
with the electron beams 
first fully right polarized and 
then fully left polarized,
we obtain the following quartic expressions 
for the highest possible resolution
obtainable with an integrated luminosity $\cal L$
\bea
\chi_\infty^2(e^+e^-) & \simeq &
{\cal L} \quad {\pi\alpha^2s \over m_{Z'}^4} \quad
\left(
  10 v_{Z'}^4 + 22 v_{Z'}^2a_{Z'}^2 + 7 a_{Z'}^2
\right)
\label{pecr}\\
\chi_\infty^2(e^-e^-) & = &
{\cal L} \quad {\pi\alpha^2s \over m_{Z'}^4} \quad
\left(
  16 R_{Z'}^4 + 16 L_{Z'}^4
\right)
\label{eecr}\\\nonumber
& = &
{\cal L} \quad {\pi\alpha^2s \over m_{Z'}^4} \quad
\left(
  32 v_{Z'}^4 + 192 v_{Z'}^2a_{Z'}^2 + 32 a_{Z'}^2
\right)
\eea

These Cramer-Rao minimum variance bounds
are the best results one can hope to ever achieve
in the absence of systematic errors 
for a given 
luminosity, 
\cm\ energy and
$Z'$ mass.
They may therefore serve as a benchmark of each reaction,
and clearly indicate that
the $e^-e^-$ experiment can be more sensitive 
to small values of $v_{Z'}$ and $a_{Z'}$,
especially in the diagonal (chiral) directions.

To be more realistic
we impose from now on a cut of $10^o$ on the scattering angle
and consider 90\%\ polarized electron beams
({\em i.e.},
a 5\%\ contamination of electrons with opposite helicity).
Moreover,
we assume the $e^-e^-$ mode
can only achieve half the luminosity of the $e^+e^-$ mode.
The relevant parameters 
at LEP2 and SLC$e^-e^-$
as well as at FLC$e^\pm e^-$
are summarized in Tables~\ref{tslc} and \ref{tnlc}.

Since the $e^-e^-$ reaction
provides very large statistics,
it is important to reduce as much as possible the systematic errors.
This can be done
by normalizing the event numbers in each bin
with respect to the total number of events.
This way,
we eliminate
the by far dominant systematic error,
which originates from the luminosity monitoring\footnote{
  Note that in M\o ller scattering
  there is no systematic error due to electron-pion misidentification,
  as in Bhabha scattering.
}.
The information from the total rates is then lost,
but we already noticed from Fig.~\ref{fxs}
that this observable is anyway not particularly relevant
for the $Z'$ signal in $e^-e^-$ scattering.

The $e^+e^-$ total \xs,
on the other hand,
is rather sensitive to the presence of a $Z'$
and we should therefore keep this observable in the analysis.
This goes without problem,
because the event rates are sufficiently low
to prevent the systematic error,
which is mainly due to the luminosity measurement,
to become dominant.

To set realistic observability limits on the $Z'$ parameters
we use the following two least squares estimators:
\beq
\chi^2_{e^+e^-} = \sum_{\rm bins}
\left(
  \frac{n^{Z'}_i-n^{SM}_i}{\Delta n^{SM}_i}
\right)^2
\qquad\qquad
\left(\Delta n_i\right)^2 = n_i + \left(\epsilon_{\rm syst} n_i\right)^2
\label{chi2pe}
\eeq

\beq
\chi^2_{e^-e^-} = \sum_{\rm bins}
\left(
  \frac{n^{Z'}_i/n^{Z'}-n^{SM}_i/n^{SM}}{\Delta n^{SM}_i/n^{SM}}
\right)^2
\qquad
\left(\Delta n_i/n\right)^2 = {n_i \over n^2} \left( 1-{n_i \over n} \right)
\label{chi2ee}~,
\eeq
where we assume a systematic error $\epsilon_{\rm syst} = 1\%$
on the $e^+e^-$ luminosity measurement.
Since the lepton angles can be measured to an accuracy better than 10 mrad,
there is no systematic error 
due to bin-to-bin correlations,
if we perfom the analysis with 10 bins. 
This way each bin is also guaranteed to contain a large number of events,
which is almost gaussian distributed.

In Fig.~\ref{fslc} we plot 
as a function of the mass of the $Z'$,
the smallest values of the reduced couplings\footnote{
  We plot the couplings divided by the $Z'$ mass 
  only for rendering the asymptotic behaviour obvious.
  We also multiply by $\sqrt{s}$ 
  to keep quantities dimensionless.
}
$R_{Z'} \sqrt{s} / m_{Z'}$ and $v_{Z'} \sqrt{s} / m_{Z'}$,
which can be detected
with 95\%\ confidence 
($\chi^2=6$)
at LEP2 or the SLC run in the $e^-e^-$ mode.
All collider parameters
are summarized in Table~\ref{tslc}.
There is of course a $Z'$ peak in the LEP2 $e^+e^-$ reaction 
for $m_{Z'} = \sqrt{s} = 180$ GeV.
Obviously,
close to this pole
the $e^+e^-$ mode provides the best information.
For higher $Z'$ masses
and chiral couplings,
the SLC$e^-e^-$
can be more sensitive than LEP2.

Similarly,
we plot in Fig.~\ref{fmzp} 
the same smallest values of the reduced couplings
as a function of the mass of the $Z'$,
for the $e^\pm e^-$ modes of a 500 GeV \lc.
The other collider parameters
are summarized in Table~\ref{tnlc}.
As anticipated,
for $Z'$ masses exceeding slightly the \cm\ energy, 
the $e^-e^-$ mode is better
(even with only half the luminosity),
especially for chiral couplings.
As it turns out,
these realistic predictions
come within only a few percent of the 
of the theoretical limits (\ref{pecr},\ref{eecr}) of each experiment.

\section{Resolving Power}

If a $Z'$ is indeed discovered,
it becomes interesting to measure as accurately as possible 
its couplings and mass.
The same procedure as in the previous section
can be applied for the parameter determination,
replacing the \sm\ rates labeled $SM$ in Eqs~(\ref{chi2pe},\ref{chi2ee}),
by the ones expected for the true values of the parameters,
and computing the $\chi^2$ variations around these values.
This is what we have done in Fig.~\ref{fresol},
at a 500 GeV collider
for a $Z'$ mass of 2 TeV and 
for several possible values of the $Z'$ vector and axial couplings.

For purely vector or axial $Z'$ couplings,
the $e^-e^-$ reaction provides a better local resolution.
For chiral couplings, though,
both processes provide complementary constraints.
This is mainly due to the fact that the $e^-e^-$ mode
cannot distinguish the four combinations of couplings 
related by the mirror symmetries around the diagonal axis
$\left( R_{Z'} \leftrightarrow -R_{Z'} \right)$
or
$\left( L_{Z'} \leftrightarrow -L_{Z'} \right)$.
The $e^+e^-$ reaction 
has only a two-fold ambiguity,
relating combinations of couplings 
located opposite of the origin
$\left( v_{Z'},a_{Z'} \leftrightarrow -v_{Z'},-a_{Z'} \right)$.
The use of $LR$ polarized beams
does not improve the $e^-e^-$ resolution 
beyond what can be achieved with $e^+e^-$.

To gauge the overall capabilities of \lc,
we have overlayed in Fig.~\ref{fmodels}
the expectations of the $Z'$ models 
described in the second section
with several 95\%\ confidence contours.
The latter have been obtained 
from the combined results of the \pe\ and \ee\ experiments,
assuming a $Z'$ mass of 1 TeV
and ignoring the ambiguity with respect to the central symmetry.
For this $Z'$ mass
the picked models are thus all well distinguishable from each other
and from the \sm.

As we can gather from Fig.~\ref{fmzp}
and Eqs~(\ref{pecr},\ref{eecr}),
if the $Z'$ is heavy compared to the collider energy,
the couplings and the mass 
are correlated in such a way that it is difficult to disentangle
a large mass from small couplings
and {\em vice versa}.
Asymptotically,
for $m_{Z'}^2 \gg s$,
the following scaling law applies
\beq
{\eta_{Z'} \over m_{Z'}}
~
\left( s {\cal L} \right)^{1\over4}
~=~ \mbox{constant}
\qquad\qquad
\eta_{Z'} = v_{Z'},a_{Z'},R_{Z'},L_{Z'}
\label{scale}
\eeq
Having determined bounds on the couplings
for a given energy, luminosity and $Z'$ mass,
it is trivial to use this scaling law
to determine the corresponding bounds on the couplings
for different energies, luminosities and $Z'$ masses.

\section{Radiative Corrections}

We still have to address the issue of radiative corrections.
They have been computed for the $e^+e^-$ case~\cite{l}
and turn out to degrade the $Z'$ bounds by about 10\%.
This is mainly due to initial state radiation
which reduces the effective \cm\ energy.
Indeed,
since the sensitivity of the $e^+e^-$ reaction 
originates from probing the tail of the $Z'$ peak,
any reduction in energy lowers the signal to background ratio.
This is clearly demonstrated by Figs~\ref{fxs},\ref{fmzp}.

In contrast,
the $e^-e^-$ process
should not be much affected,
because it probes the angular distribution,
which is not significantly modified by radiative corrections
in the considered angular range.

\section{Conclusions}

We have performed a comparison of $Z'$ searches 
in the $e^+e^-$ and $e^-e^-$ modes of a \lc\ of the next generation.
For this,
we used a similar analysis for each process
and assumed the same collider parameters,
except for the $e^-e^-$ luminosity
which we set to only half the $e^+e^-$ luminosity.

Both analysis procedures are rather efficient,
yielding results very close to the theoretical Cramer-Rao limit.
The $e^-e^-$ mode turns out to be slightly better 
for discovering a $Z'$,
while both reactions are complementary 
when it comes to determining the $Z'$-lepton couplings.

For completeness,
we have also compared the discovery potential of LEP2 {\em vs} SLC\ee.
Here also,
both machines yield similar bounds,
LEP2 being more performant for a light $Z'$ 
and SLC\ee\ for a heavy $Z'$.

\clearpage

\begin{table}[htb]
  \begin{center}
    \begin{tabular}{|l|cc|cc|}
      \hline
      collider & \multicolumn{2}{c|}{LEP2} & \multicolumn{2}{c|}{SLC$e^-e^-$} \\
      \hline
      beams & $e^+$ & $e^-$ & $e^-$ & $e^-$ \\
      beam polarization [\%] & 0 & 0 & 90 & 90 \\
      energy [GeV] & \multicolumn{2}{c|}{180} & \multicolumn{2}{c|}{100} \\
      luminosity [pb$^{-1}$] & \multicolumn{2}{c|}{$10$} & \multicolumn{2}{c|}{$2\times1$} \\
      polar angle cut [$^o$] & \multicolumn{2}{c|}{10} & \multicolumn{2}{c|}{10} \\
      number of bins & \multicolumn{2}{c|}{10} & \multicolumn{2}{c|}{10} \\
      systematic error [\%] & \multicolumn{2}{c|}{1} & \multicolumn{2}{c|}{0} \\
      \hline
    \end{tabular}
  \end{center}
  \caption{
    Comparison of the different parameters 
    used in the LEP2 and SLC$e^-e^-$ analysis.
    }
  \label{tslc}
\end{table}

\vfill

\begin{table}[htb]
  \begin{center}
    \begin{tabular}{|l|cc|cc|}
      \hline
      collider & \multicolumn{2}{c|}{FLC$e^+e^-$} & \multicolumn{2}{c|}{FLC$e^-e^-$} \\
      \hline
      beams & $e^+$ & $e^-$ & $e^-$ & $e^-$ \\
      beam polarization [\%] & 0 & 90 & 90 & 90 \\
      energy [GeV] & \multicolumn{2}{c|}{500} & \multicolumn{2}{c|}{500} \\
      luminosity [fb$^{-1}$] & \multicolumn{2}{c|}{$2\times20$} & \multicolumn{2}{c|}{$2\times10$} \\
      polar angle cut [$^o$] & \multicolumn{2}{c|}{10} & \multicolumn{2}{c|}{10} \\
      number of bins & \multicolumn{2}{c|}{10} & \multicolumn{2}{c|}{10} \\
      systematic error [\%] & \multicolumn{2}{c|}{1} & \multicolumn{2}{c|}{0} \\
      \hline
    \end{tabular}
  \end{center}
  \caption{
    Comparison of the different parameters 
    used in the FLC$e^\pm e^-$ analysis.
    }
  \label{tnlc}
\end{table}

\vfill

\begin{figure}[htb]
\unitlength.5mm
\SetScale{1.418}
\begin{boldmath}
\begin{center}
\begin{picture}(80,40)(0,0)
\ArrowLine(0,0)(15,15)
\ArrowLine(15,15)(0,30)
\Photon(15,15)(45,15){2}{5}
\ArrowLine(60,30)(45,15)
\ArrowLine(45,15)(60,0)
\Text(-2,0)[r]{$e^-$}
\Text(-2,30)[r]{$e^+$}
\Text(62,0)[l]{$\mu^-$}
\Text(62,30)[l]{$\mu^+$}
\Text(30,23)[c]{$\gamma$,${Z^0}$,$Z'$}
\end{picture}
\qquad\qquad\qquad
\begin{picture}(80,40)(0,0)
\ArrowLine(0,0)(30,0)
\ArrowLine(30,0)(60,0)
\ArrowLine(0,30)(30,30)
\ArrowLine(30,30)(60,30)
\Photon(30,0)(30,30){2}{5}
\Text(-2,0)[r]{$e^-$}
\Text(-2,30)[r]{$e^-$}
\Text(62,0)[l]{$e^-$}
\Text(62,30)[l]{$e^-$}
\Text(34,15)[l]{$\gamma$,${Z^0}$,$Z'$}
\end{picture}
\end{center}
\end{boldmath}
\caption{
  Lowest order Feynman diagrams for the neutral gauge boson exchanges 
  in $e^+e^-$ and $e^-e^-$ scattering.
}
\label{feyn}
\end{figure}
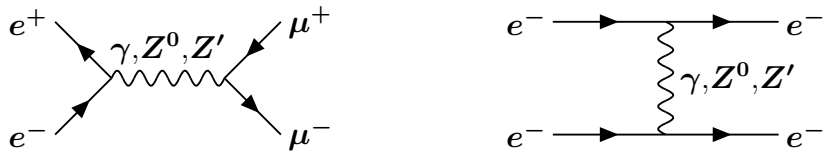

\clearpage

\begin{figure}[htb]
\begin{boldmath}
\begin{center}
\input{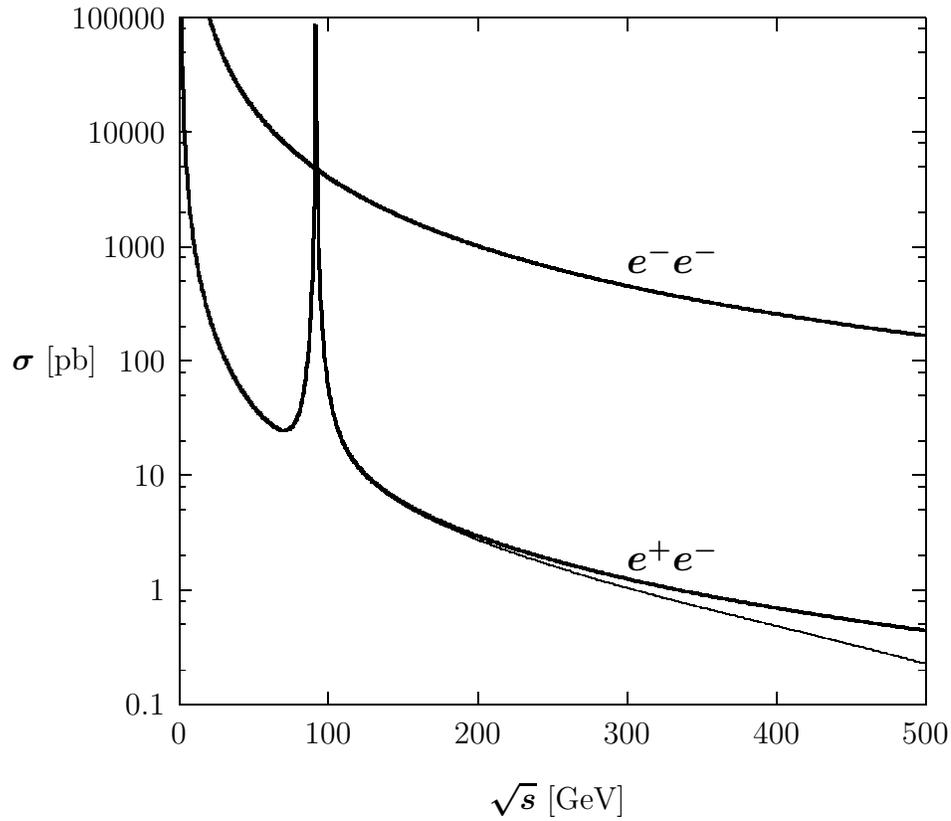}
\end{center}
\end{boldmath}
%\vspace{-1mm}
\caption{
  Energy dependence of the unpolarized $e^\pm e^-$ \xs s 
  in the framework of the \sm\ (thick curves)
  and with a 1 TeV $Z'$ which couples like a photon (thin curves).
  In the $e^-e^-$ case
  the two curves cannot be resolved on this scale.
}
\label{fxs}
\end{figure}

\clearpage

\begin{figure}[htb]
\begin{boldmath}
\begin{center}
\input{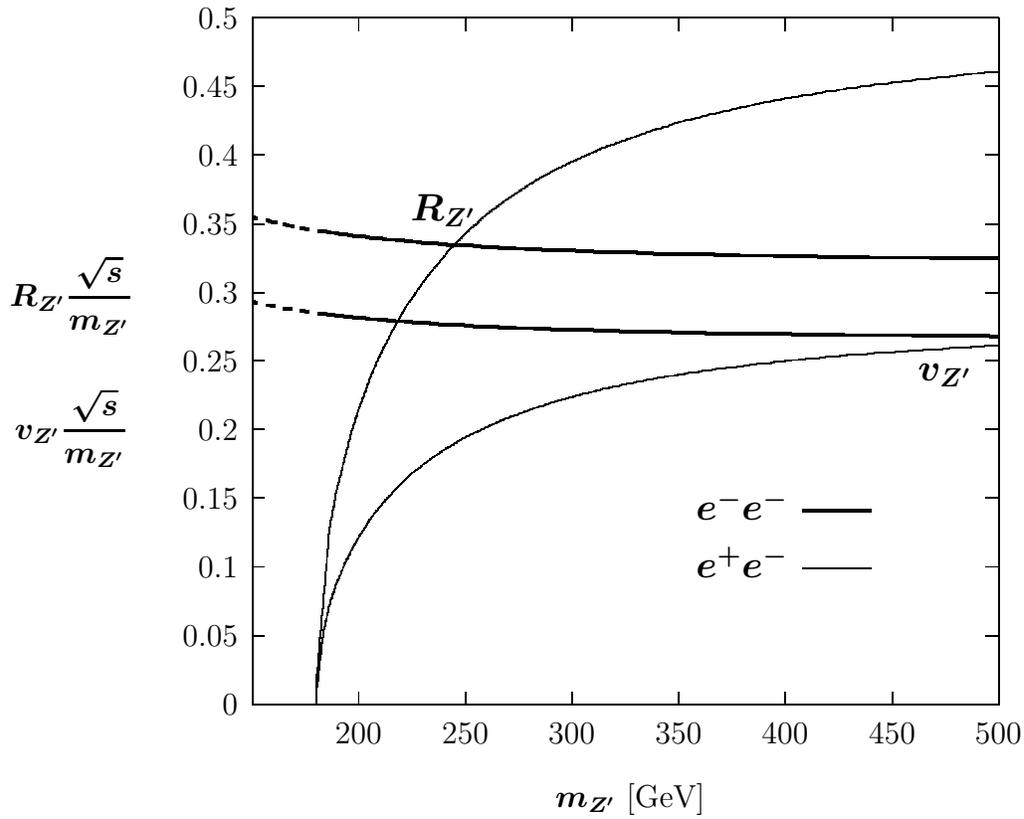}
\end{center}
\end{boldmath}
\caption{
  Smallest measurable values of normalized $Z'$ couplings 
  with 95\%\ confidence
  as functions of the $Z'$ mass,
  at LEP2 and SLC\ee
  ({\em cf.} Table~\protect\ref{tslc}).
}
\label{fslc}
\end{figure}

\clearpage

\begin{figure}[htb]
\begin{boldmath}
\begin{center}
\input{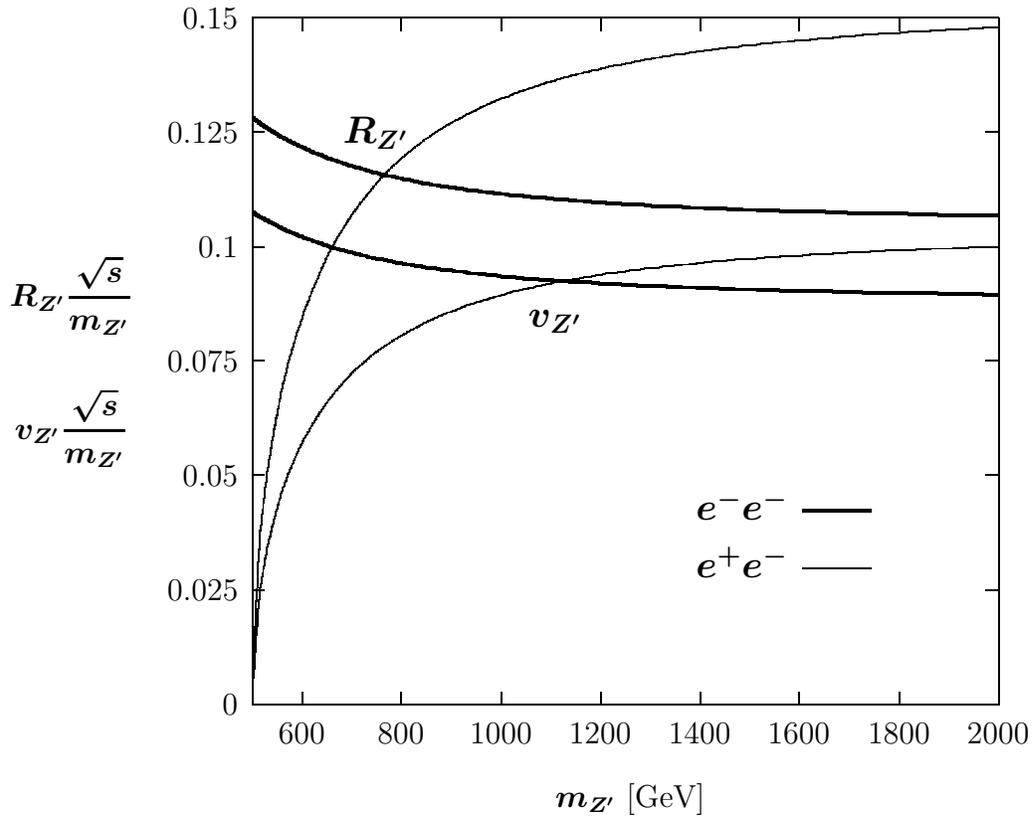}
\end{center}
\end{boldmath}
\caption{
  Smallest measurable values of normalized $Z'$ couplings 
  with 95\%\ confidence
  as functions of the $Z'$ mass,
  in \pe\ and \ee\ collisions
  at the FLC
  ({\em cf.} Table~\protect\ref{tnlc}).
}
\label{fmzp}
\end{figure}

\clearpage

\begin{figure}[htb]
\begin{boldmath}
\begin{center}
\input{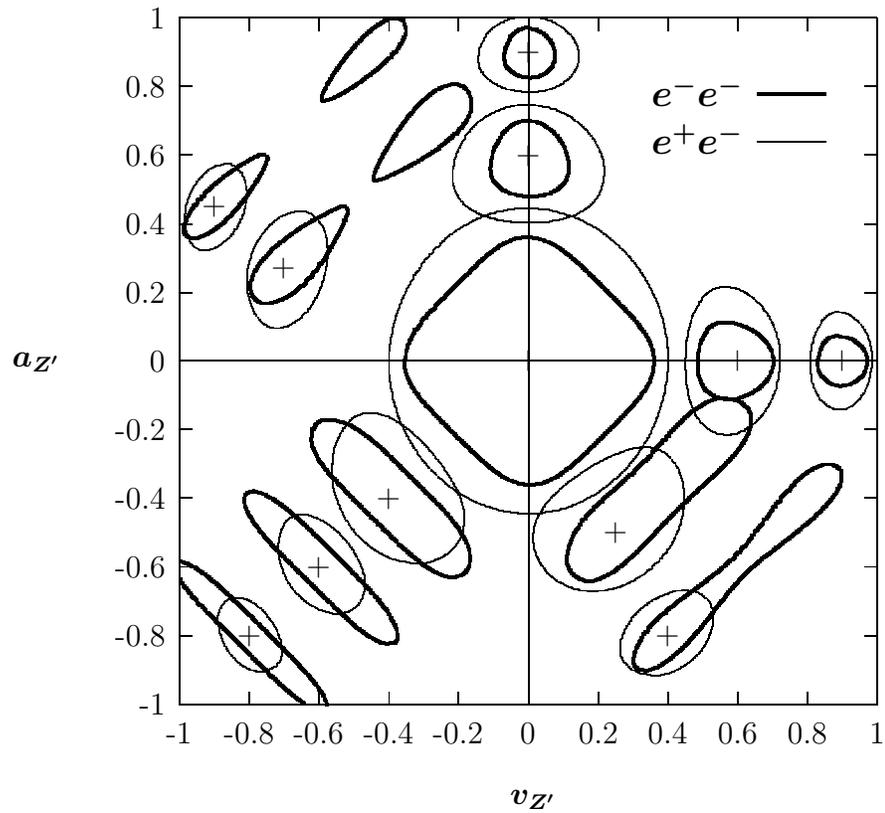}
\end{center}
\end{boldmath}
\caption{
  Contours of resolvability at 95\%\ confidence
  of the $Z'$ couplings
  around several possible true values 
  marked with a `+'.
  The FLC characteristics are summarized in Table~\protect\ref{tnlc}
  and the $Z'$ mass is 2 TeV.
}
\label{fresol}
\end{figure}

\clearpage

\begin{figure}[htb]
\begin{boldmath}
\begin{center}
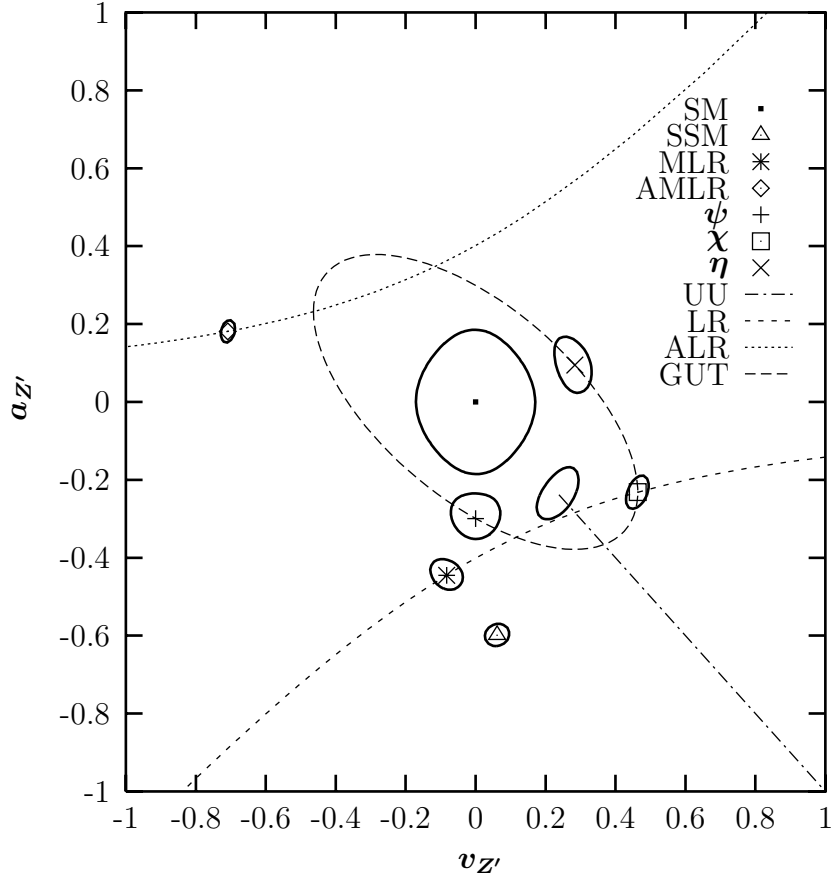
\end{center}
\end{boldmath}
\caption{
  Contours of resolvability at 95\%\ confidence
  of the $Z'$ couplings
  around several possible model predictions.
  The information from \pe\ and \ee\ collisions is combined.
  The FLC characteristics are summarized in Table~\protect\ref{tnlc}
  and the $Z'$ mass is 1 TeV.
}
\label{fmodels}
\end{figure}

\end{document}